\documentclass[aps,prl,superscriptaddress,amsfonts,graphicx,reprint]{revtex4-2}
\usepackage[utf8]{inputenc}
\usepackage[T1]{fontenc}
\usepackage{tgtermes}
\usepackage{amsmath, amssymb}
\usepackage{mathtools}
\usepackage{breqn}
\usepackage{graphicx}
\usepackage{float}
\usepackage{bm}
\usepackage{physics}
\usepackage{xcolor}
\usepackage{verbatim}
\usepackage{enumitem}

\graphicspath{{fig/}}
\usepackage{hyperref}
\usepackage{multirow,xcolor}
\hypersetup{
    colorlinks = true,
    linkcolor = blue,
    linkbordercolor = white,
    citecolor=blue,
    urlcolor=black
}
\usepackage[export]{adjustbox}
\usepackage{subfig,caption}
\newcommand{\ansatz}{\textit{ansatz }}
\newcommand{\abinitio}{\textit{ab initio }}

\makeatletter
\let\cat@comma@active\@empty
\makeatother

\begin{document}
\title{A Phase-Space Semiclassical Approach for Modeling Nonadiabatic Nuclear Dynamics with Electronic Spin}
\author{Yanze Wu}
\email{wuyanze@sas.upenn.edu}
\affiliation{Department of Chemistry, University of Pennsylvania, Philadelphia, Pennsylvania 19104, USA}
\author{Xuezhi Bian}
\email{xzbian@sas.upenn.edu}
\affiliation{Department of Chemistry, University of Pennsylvania, Philadelphia, Pennsylvania 19104, USA}
\author{Jonathan Rawlinson}
\email{jonathan.rawlinson@manchester.ac.uk}
\affiliation{School of Mathematics, The University of Manchester, Oxford Rd, Manchester M13 9PL, United Kingdom}
\author{Robert G. Littlejohn}
\email{robert@wigner.berkeley.edu}
\affiliation{Department of Physics, University of California, 366 Physics North MC 7300, Berkeley, CA, 94720-7300}
\author{Joseph E. Subotnik}
\email{subotnik@sas.upenn.edu}
\affiliation{Department of Chemistry, University of Pennsylvania, Philadelphia, Pennsylvania 19104, USA}
\date{\today}

\begin{abstract}
Chemical relaxation phenomena, including photochemistry and electron transfer processes, form a vigorous area of research in which nonadiabatic dynamics plays a fundamental role. Here, we show that for nonadiabatic dynamics with two electronic states and a complex-valued Hamiltonian that does not obey time-reversal symmetry, the optimal semiclassical approach is to run surface hopping dynamics on a set of phase-space adiabatic surfaces. In order to generate such phase-adiabats, one must isolate a proper set of diabats and apply a phase gauge transformation, before eventually diagonalizing the total Hamiltonian (which is now parameterized by both ${\bf R}$ and ${\bf P}$). The resulting algorithm is valid in both the adiabatic and nonadiabatic limits, incorporates all Berry curvature effects, and allows for the study of semiclassical nonadiabatic dynamics in the presence of spin-orbit coupling and/or external magnetic fields.
\end{abstract}

\maketitle

\paragraph{Introduction.}
Coupled nuclear-electronic, nonadiabatic dynamics underlie critical aspects of many photochemical \cite{mai2020Molecular,nelson2020Nonadiabatic,stock2005Classical,levine2007Isomerization,penfold2018SpinVibronic} and electron transfer processes \cite{closs1988Determination,scholes2017Using}. The basic premise is that, when electronic transitions occur, energy must be provided or absorbed by the nuclei, 
and there are a host\cite{stock2005Classical} of standard approaches for modeling such nonadiabatic energy conversion, including Ehrenfest dynamics \cite{mclachlan1964variational}, surface hopping \cite{tully1990Molecular}, multiple spawning \cite{ben-nun2000Initio} and exact factorization \cite{min2015Coupledtrajectory}.
Although not usually considered within the chemical physics community, nonadiabatic effects can also arise that conserve energy within the context of molecular dynamics; i.e., nonadiabatic effects can arise that bend nuclear trajectories without changing their kinetic energy. For instance, single surface on-diagonal Berry curvature effects can arise when there is an external magnetic field and the Hamiltonian is complex-valued. \cite{berry1993Chaotic,takatsuka2007Generalization,subotnik2019demonstration,culpitt2021initio}. In such a case, the nuclei experience a Lorentz-like force on their motion. In the adiabatic limit, this force is \cite{berry1993Chaotic}
\begin{align}
    \mathbf{F}^B_n = i\hbar\dot{\mathbf{R}}\times(\nabla\times\mathbf{D}_{nn}^A)
    \label{eq:bf}
\end{align}
where $n$ is the adiabatic surface, $\dot{\mathbf{R}}$ is the nuclear velocity and $\mathbf{D}_{nn}^A$ is the derivative coupling (also called Berry connection) on surface $n$. More generally, one can argue that nonadiabatic pseudo-magnetic field effects occur whenever there are degenerate or nearly degenerate electronic states coupled together, e.g. when one considers spin states coupled together with spin-orbit coupling.\cite{bian2022Incorporating} These effects may well explain 
many cutting-edge spin-related chemical and physical reactions, including chiral induced spin selectivity (CISS) \cite{naaman2012ChiralInduced,fransson2020Vibrational} or other magnetic chemical reactions \cite{hore2020Spina}.


In order to better understand how nonadiabatic dynamics, Berry curvature and the presence of spins does or does not affect chemical dynamics, it is essential to have cheap, inexpensive semi-classical algorithms. A proper algorithm must capture both the magnitude of a momentum change upon hopping (in the spirit of Tully's trajectory surface hopping \cite{tully1990Molecular}) and the pseudo-magnetic Berry force that rotates momentum (in the spirit of Berry's half-classical dynamics \cite{berry1993Chaotic}); to date, there is no well established, reliable protocol. 
For instance, the simplest test case is a system with an odd number of electrons, two accessible spatial electronic states, and two possible spin states; the total Hilbert space is four-dimensional. If one ignores spin flips, one can separate this $4 \times 4$ Hamiltonian into a pair of $2 \times 2$ complex-valued Hamiltonians.\cite{chandran2022Electron} 
Previously, we have made several attempts to study such $2 \times 2$ complex-valued Hamiltonians by incorporating the Berry curvature effect with Tully's fewest switch surface hopping (FSSH) \cite{miao2019extension,miao2020backtracking,wu2021Semiclassical,bian2022Incorporating}, but the final algorithm has inevitably
failed when the nonadiabatic effects became strong enough \cite{miao2019extension,wu2021Semiclassical}.

With this failure in mind, below we show that the solution is to run semiclassical phase-space surface hopping (PSSH) calculations in the spirit of (but not equivalent to) Ref. \cite{shenvi2009Phasespace}. According to PSSH, trajectories move on phase-space adiabatic surfaces $E(\mathbf{R},\mathbf{P})$ which are functions of both nuclear position and momentum. 
For a two-state problem, the PSSH approach effectively transforms a complex-valued Hamiltonian into a real-valued Hamiltonian, all while preserving the basic avoided crossing structure; the present PSSH algorithm also conveniently reduces to standard FSSH for a real-valued Hamiltonian. 
Compared to our previous work on FSSH extensions \cite{miao2019extension,wu2021Semiclassical}, the present pseudo-diabatic PSSH is simpler, more accurate, and more general. The present approach is also applicable for modeling dynamics in a magnetic field or under illumination by circularly polarized light.\cite{schellman1975Circular}

\begin{figure}
    \begin{center}
    \subfloat{\includegraphics[width=0.8\columnwidth]{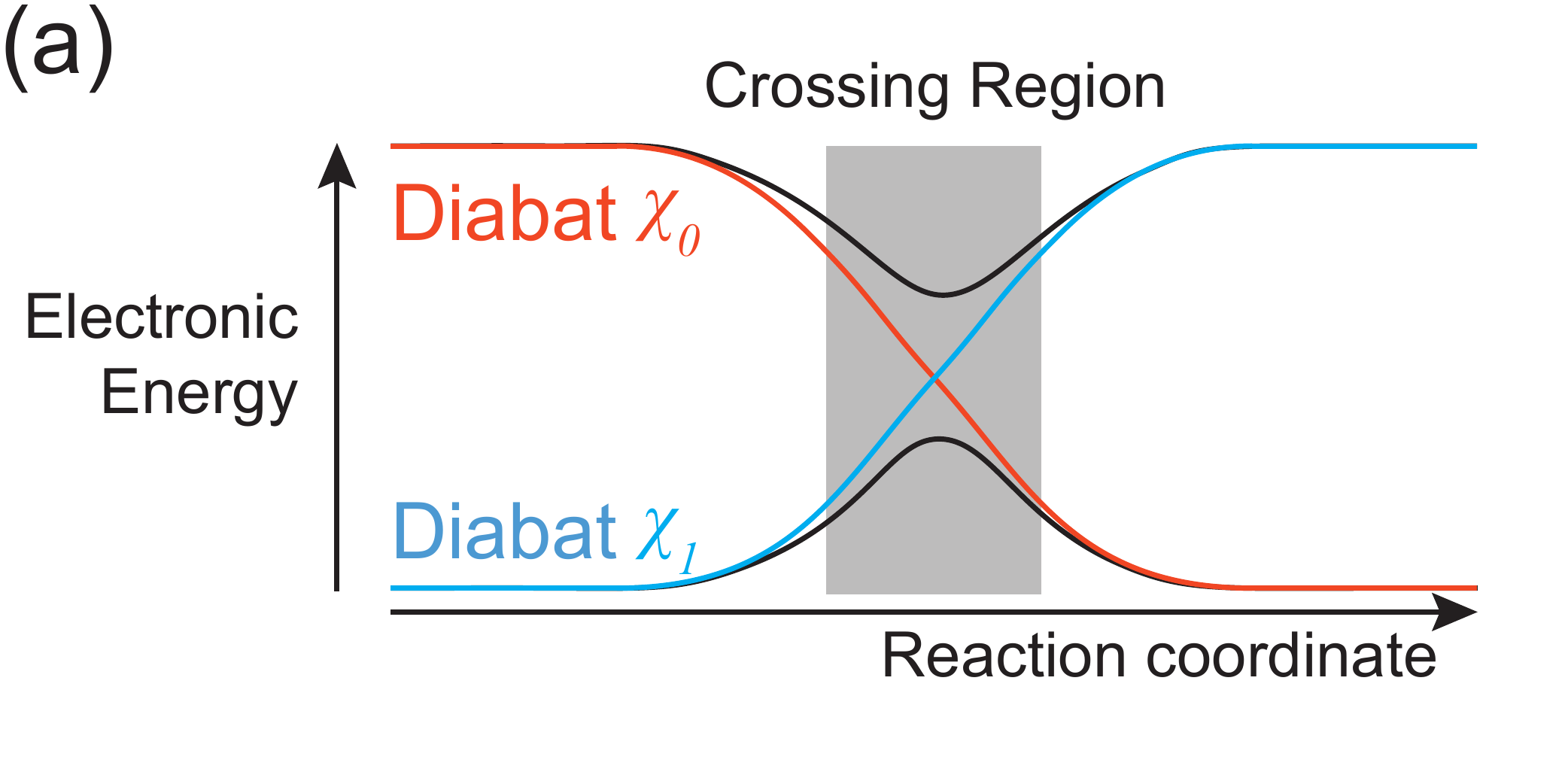} \label{fig:crossing}}
    
    \subfloat{\includegraphics[width=0.9\columnwidth]{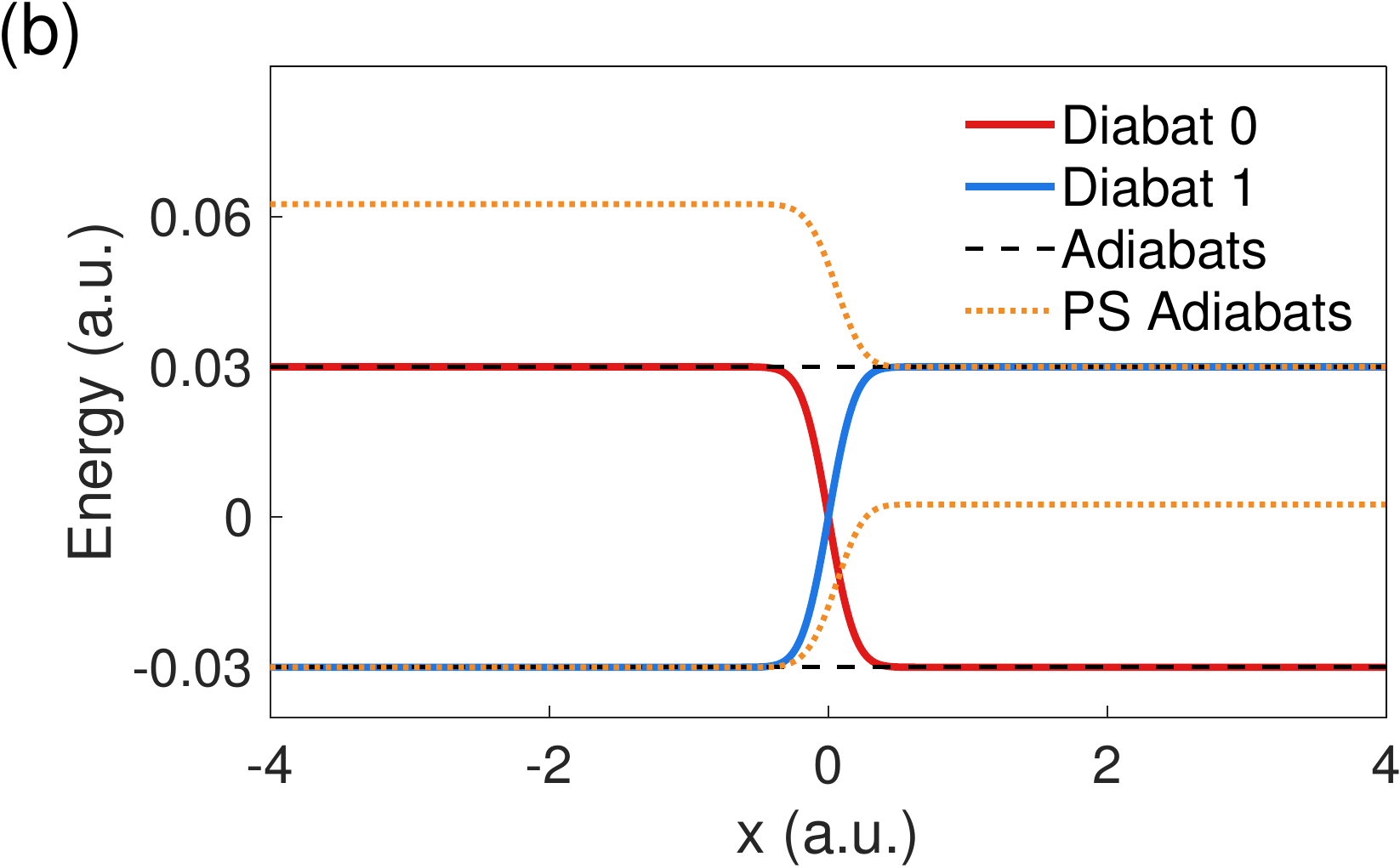} \label{fig:model}}
    \end{center}
    \caption{(a) A schematic depiction of a curve crossing, where $\chi_0$ and $\chi_1$ are two proper diabats. (b) The diabatic and (position-space) adiabatic surfaces of our test model (see Eq.~\eqref{eq:hmodel}) as well as typical phase-space adiabatic surfaces (shifted by $-P^2/2M$), as functions of nuclear coordinate $x$. Note that the position-space adiabats are flat, while the phase-space adiabats have a barrier, a distinct signature of complex-valued Hamiltonians. The parameters used to plot the phase-space adiabats are $W=-5$, $P_y=8$ and $\chi_{\text{init}}=\chi_0$. }
\end{figure}

\paragraph{Construction of the Phase-Space Hamiltonian.}
Consider a general two-state nonadiabatic Hamiltonian:
\begin{align}
    \hat{H} &= \frac{\hat{\mathbf{P}}^2}{2M} + \hat{h}_{\text{el}}(\hat{\mathbf{R}},\hat{\mathbf{r}}) \label{eq:fullh}
\end{align}
where $\hat{\mathbf{P}}$ and $\hat{\mathbf{R}}$ are the nuclear momentum and position operators and $\hat{\mathbf{r}}$ represents the electronic degrees of freedom.
A common situation is an avoided crossing. The typical topology of an avoided crossing is shown in Fig.~\ref{fig:crossing}: the two diabats cross each other, and the adiabats are repelled by the diabatic couplings.
For this Letter, we will focus on a very simple avoided crossing. First, we assume that the pair of states cross only once. Second, we assume there is a pair of ``proper diabats'' that coincides with the adiabats asymptotically, just as shown in Fig.~\ref{fig:crossing}. Based on these two assumptions, we can write our electronic Hamiltonian in the proper diabatic basis $\ket{\chi_0}$ and $\ket{\chi_1}$ in the vicinity of the crossing as
\begin{align}
    \hat{h}_{\text{el}} &= \begin{bmatrix} h_0 (\hat{\mathbf{R}}) & V(\hat{\mathbf{R}})e^{i\phi(\hat{\mathbf{R}})} \\ V(\hat{\mathbf{R}})e^{-i\phi(\hat{\mathbf{R}})} & h_1(\hat{\mathbf{R}}) \end{bmatrix} \label{eq:hel}
\end{align}
where the proper diabatization requires $\abs{V}\ll\abs{h_0-h_1}$ outside the crossing seam. This Hamiltonian does not obey time reversal symmetry.

Within the usual Born-Oppenheimer picture, one rotates Hamiltonian~\eqref{eq:fullh} to the adiabatic basis, where the nuclear motion is coupled to electronic amplitudes via the derivative coupling terms \cite{cederbaum2004Bornoppenheimer}. However, here we will make a different choice: we will represent Hamiltonian~\eqref{eq:fullh} in a pseudo-diabatic basis $\ket{\xi_0}=\ket{\chi_0}$, $\ket{\xi_1}=e^{-i\phi}\ket{\chi_1}$ where we assign phases but not rotations to a set of diabats. The result is a pseudo Born-Oppenheimer Hamiltonian:
\begin{align}
    \hat{H}_{\text{PD}} = \frac{{(\hat{\mathbf{P}} - i\hbar\hat{\mathbf{D}})}^2}{2M} + \begin{bmatrix}h_0(\hat{\mathbf{R}})& V(\hat{\mathbf{R}}) \\ V(\hat{\mathbf{R}}) & h_1(\hat{\mathbf{R}}) \end{bmatrix}
    \label{eq:pdfullh}
\end{align}
where $\hat{\mathbf{D}}=-i\nabla\phi\ket{\xi_1}\bra{\xi_1}$ is the derivative coupling in this pseudo-diabatic basis. Note here that $\hat{\mathbf{iD}}$, $h_0$, $h_1$ and $V$ are all real-valued; by performing a pseudo-diabatic transformation, we have turned the complex-valued Hamiltonian~\eqref{eq:fullh} into a real-valued Hamiltonian~\eqref{eq:pdfullh}, which will enable us to use simple (or simpler) semiclassical approaches for modeling.
For a deeper discussion of the choice pseudo-diabats, see the SI.

To implement semiclassical (surface-hopping) dynamics, we first replace the nuclear operators in Hamiltonian~\eqref{eq:pdfullh} by their classical counterparts (in the spirit of a Wigner transformation\cite{kapral1999Mixeda}):
\begin{align}
    H_{\text{PD}}(\mathbf{R},\mathbf{P}) = \frac{{(\mathbf{P} - i\hbar\mathbf{D}(\mathbf{R}))}^2}{2M} + \begin{bmatrix}h_0(\mathbf{R})& V(\mathbf{R}) \\ V(\mathbf{R}) & h_1(\mathbf{R}) \end{bmatrix} 
    \label{eq:pdhclassical}
\end{align}
Second, after diagonalizing Hamiltonian~\eqref{eq:pdhclassical}, we arrive at a basis depending on both position $\mathbf{R}$ and momentum $\mathbf{P}$: 
\begin{align}
    H_{\text{PD}}(\mathbf{R},\mathbf{P})\ket{\psi_j(\mathbf{R},\mathbf{P})}=E_j(\mathbf{R},\mathbf{P})\ket{\psi_j(\mathbf{R},\mathbf{P})}
    \label{eq:superad}
\end{align}
We will call the resulting eigenvalues and eigenvectors ``phase-space adiabats.''

In some sense, this new basis mimics what Berry has labeled ``superadiabats'' \cite{berry1987Quantum,berry1990Histories}, i.e. the basis recovered by first diagonalizing the electronic Hamiltonian $h_{\text{el}}(\mathbf{R})$ and then second re-diagonalizing the sum of adiabatic electronic energies $E_A(\mathbf{R})$, the kinetic term and the relevant derivative couplings $\mathbf{D}_A$ \cite{berry1987Quantum,berry1990Histories,takatsuka2006NonBornOppenheimer,shenvi2009Phasespace}:
\begin{align}
H_{\text{super}}(\mathbf{R},\mathbf{P}) = \frac{{(\mathbf{P}-i\hbar\mathbf{D}_A(\mathbf{R}))}^2}{2M}+\begin{bmatrix} E_0^A(\mathbf{R}) & 0 \\ 0 & E_1^A(\mathbf{R}) \end{bmatrix}
\end{align}
Interestingly, Shenvi proposed phase-space surface-hopping dynamics more than ten years ago (for real-valued Hamiltonians) and the idea has some clear benefits (and a few problems).\cite{shenvi2009Phasespace,gherib2016inclusion}
That being said, we must be clear that the present basis $\left\{\ket{\psi}\right\}$ is not exactly the same as the superadiabatic basis; the $\left\{\ket{\xi}\right\}$'s are pseudo-diabats and not adiabats. In fact, for a real-valued Hamiltonian where $\phi\equiv 0$, $\mathbf{D}$ will always be 0 and the basis set $\left\{\ket{\psi} \right\}$ is identical to the usual position-space adiabats. Thus, though certainly related, for clarity, one should not confuse the concept of a superadiabat and the concept of a phase-space adiabat; one must also distinguish between Shenvi's adiabatic PSSH algorithm and the present pseudo-diabatic PSSH algorithm.

\paragraph{Phase-space Surface Hopping.}
Following Shenvi \cite{shenvi2009Phasespace} in spirit, we now propose to propagate the semiclassical dynamics by moving nuclei along phase-space eignvalues and then allowing for surface hops. At the beginning of the simulation, we initialize a swarm of trajectories, each associated with an electronic amplitude vector $\mathbf{c}$ and an active phase-space adiabatic label $n$. Note that the phase-space momentum $\mathbf{P}$ is different from the kinetic momentum $\mathbf{P}_{\text{kinetic}}=M\dot{\mathbf{R}}$ in general, and should be transformed according to
\begin{align}
    \mathbf{P}_n = \mathbf{P}_{\text{kinetic}} + i\hbar \mel{\psi_n}{\mathbf{D}}{\psi_n}
    \label{eq:momentum}
\end{align}
before the simulation begins.

At each time step of the simulation, we construct Hamiltonian~\eqref{eq:pdhclassical} and diagonalize it according to Eq.~\eqref{eq:superad} for each trajectory. The trajectory's equation of motion is then given by
\begin{align}
    \dot{\mathbf{R}} &= \mathbf{\nabla}_P{E_n} \\
    \dot{\mathbf{P}} &= -\mathbf{\nabla}_R{E_n} \\
    \dot{c}_j &= -\frac{i}{\hbar}E_jc_j - \mathbf{d}^R_{jk}\cdot\dot{\mathbf{R}}c_k - \mathbf{d}^P_{jk}\cdot\dot{\mathbf{P}}c_k
\end{align}
where $\mathbf{d}^R_{jk}=\braket{\psi_j}{\mathbf{\nabla}_R\psi_k}$ and $\mathbf{d}^P_{jk}=\braket{\psi_j}{\mathbf{\nabla}_P\psi_k}$ are the phase-space analogs of the derivative couplings. Note that the dynamics above conserve the energy of the relevant phase-space adiabat, i.e. $\mathrm{d}{E_n}/\mathrm{d}t=0$ along any given trajectory. 

Similar to FSSH, within PSSH, trajectories are allowed to change their active phase-space adiabatic label, or `hop' between phase-space adiabats at each step. The hopping probability from surface $k$ to $j$ is computed according to  Tully's method \cite{tully1990Molecular,shenvi2009Phasespace}:
\begin{align}
    g_{k\to j} &= \frac{\dot{\rho}_{jj}\Delta t}{\rho_{kk}} \nonumber \\ &= \frac{2\Delta t}{\hbar}\Im{\frac{c_j^*}{c_k^*}\Big(-i\hbar\mathbf{d}^R_{jk}\cdot\dot{\mathbf{R}} - i\hbar\mathbf{d}^P_{jk}\cdot\dot{\mathbf{P}}\Big) }
    \label{eq:hop}
\end{align}
Whenever a hop from $j\to k$ succeeds, we rescale the momentum along the direction of $\mathbf{d}_{jk}^R$ (which is real-valued by construction) to conserve energy. If such momentum cannot be found, the hop is frustrated and the trajectory keeps moving along the original surface. 

Finally, to capture the decoherence of a reflected wavepacket, we further employ the most naive decoherence algorithm possible, similar to what was published in Ref.~\cite{wu2021Semiclassical}, i.e. we collapse the amplitudes by setting $c_j \rightarrow \delta_{nj}$ if we find $(\mathbf{P}\cdot\mathbf{d}_{nj}^R) (\mathbf{P}_{t=0}\cdot\mathbf{d}_{nj}^R) < 0$. Here, $n$ is the active surface. We will say more about decoherence below.

\paragraph{Computational Results.}
To test the performance our algorithm, we study the simplest (standard) two-state $\left\{ \ket{\chi_0}, \ket{\chi_1} \right\}$ 
electronic Hamiltonian associated with two nuclear degrees of freedom, $x$ and $y$:
\begin{align}
    h_{\text{el}}(x,y) = A \begin{bmatrix} -\cos{\theta} & e^{iWy}\sin{\theta} \\ e^{-iWy}\sin{\theta} & \cos{\theta} \end{bmatrix}
    \label{eq:hmodel}
\end{align}
where $\theta = \frac{\pi}{2}(\erf(Bx)+1)$, $A=0.03$, $B=3$ and $W=\pm 5$. All parameters above are in atomic units. The diabatic, (position-space) adiabatic surfaces and typical phase-space adiabatic surfaces are shown in Fig.~\ref{fig:model}. Note that the position-space adiabats are completely flat, but the phase-space adiabats are typically not. The initial wavefunction is chosen as a Gaussian:
\begin{align}
    \Psi_0(\mathbf{R}) = e^{-{(\mathbf{R}-\mathbf{R}_0)}^2/\sigma^2 + i\mathbf{P}_0\cdot\mathbf{R}}\ket{\chi_{\text{init}}}
\end{align}
where $\sigma=1$, $\mathbf{R}_0=(-3,-3)$, $\mathbf{P}_0=(P_{\text{init}},P_{\text{init}})$, and $\chi_{\text{init}}$ is either the diabat 0 or 1. To make sure that the kinetic momentum equals to the phase-space momentum at $t=0$, the pseudo-diabats $\left\{\ket{\xi_0},\ket{\xi_1}\right\}$ are chosen according to the initial diabat: If $\chi_{\text{init}}=\chi_0$, then $\ket{\xi_0} = \ket{\chi_0}$ and $\ket{\xi_1} = \ket{\chi_1}e^{-iWy}$, otherwise $\ket{\xi_1}=\ket{\chi_1}$ and $\ket{\xi_0}=\ket{\chi_0}e^{iWy}$.
The exact quantum mechanics is performed using a split-operator method \cite{kosloff1983Fourier} with a $768\times 768$ grid inside a $48\times 48$ box and a timestep of 0.05 au. The surface hopping simulations were performed with $10^4$ trajectories with a timestep of 0.05 au for each data point. The initial positions and momenta for surface hopping simulations are sampled according to the Wigner distribution of $\Psi_0(\mathbf{R})$. At each point in time, the phases of the phase-space adiabatic basis can be trivially chosen according to the ``parallel transport'' condition (i.e. $\braket{\phi_j(t)}{\phi_j(t+dt)}\approx 1$ for all $j$'s). Since the diabats and phase-space adiabats are equivalent outside the crossing, the diabatic population can be computed by counting trajectories on each phase-space surface adiabat.

\begin{figure}[H]
    \begin{center}
        \includegraphics[width=0.7\textwidth]{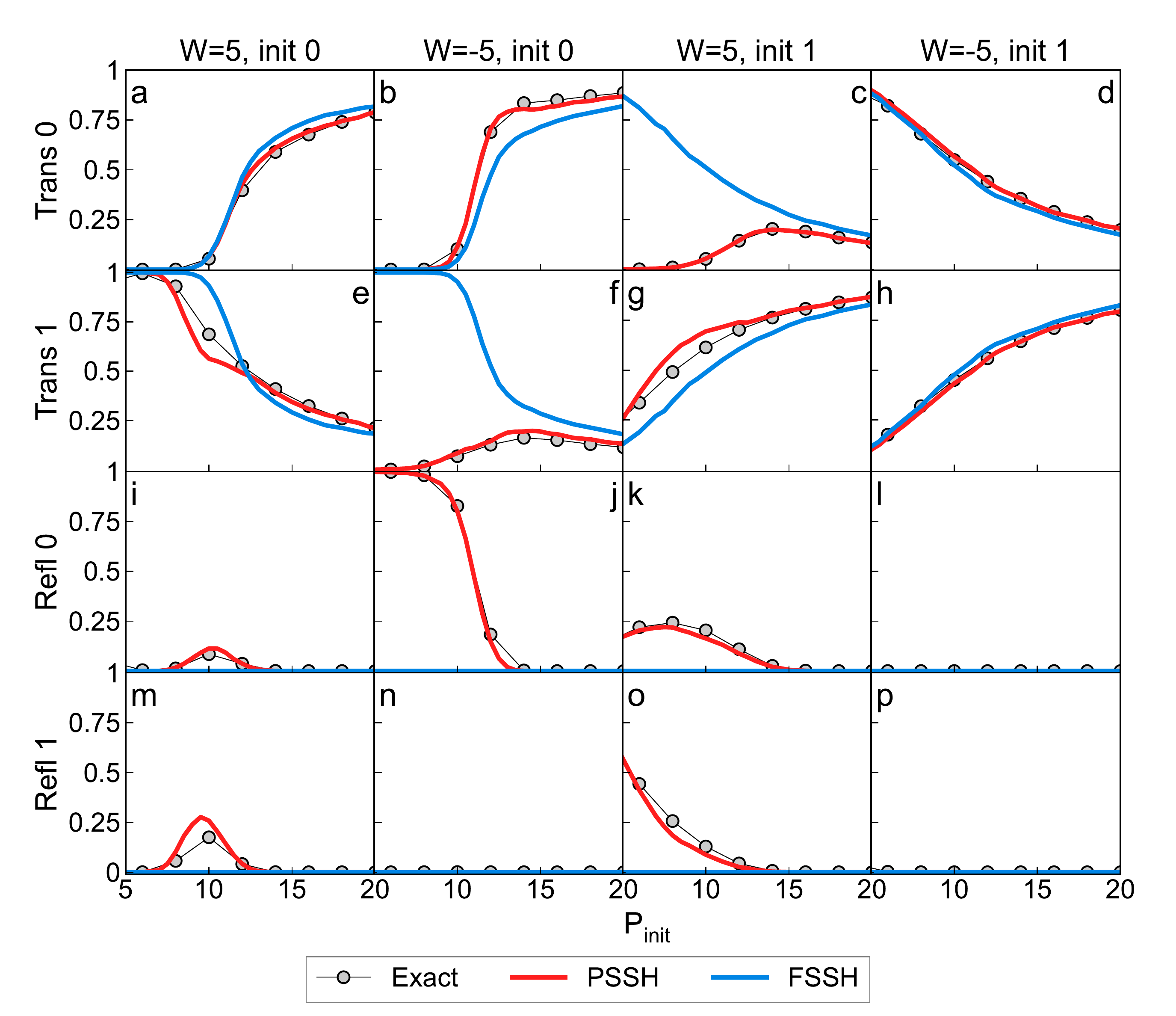}
    \end{center}
    \caption{State-to-state transmitted and reflected probabilities according to an exact wavepacket simulation, pseudo-diabatic PSSH and FSSH for our test system (Eq.~\eqref{eq:hmodel}). We have tested four conditions: $W=\pm 5$ and the initial diabat is either 0 or 1. Note that reflections are prevalent at low incoming momentum, which is a signature of Berry curvature effects. The pseudo-diabatic PSSH results agree reasonably well with the exact simulations while FSSH results deviate significantly for reflection. Parameters are: $A=0.03,B=3,M=1000$. }\label{fig:result}
\end{figure}

In Fig.~\ref{fig:result}, we compare the transmitted and reflected populations on the different surfaces according to exact wavepacket simulations, Tully's FSSH approach \cite{tully1990Molecular} and our current pseudo-diabatic PSSH simulations.
We find that in many systems, a considerable fraction of the population will be reflected when the momentum is relatively low (e.g. $P_{\text{init}}<12$). If one assumes that trajectories follow position-space adiabatic surfaces, such reflection must be a characteristic of a Berry curvature effect; after all, the forces here are completely flat. From the phase-space point of view, however, the reflection clearly arises from the barrier present in the phase-space adiabatic surfaces; see Fig. \ref{fig:model}. 
Moreover, according to Fig. \ref{fig:result}, when $W=5$ and one begins on the upper diabat, the reflected population is distributed over both diabats 0 and 1, indicating that there can be no clean separation of nonadiabatic dynamics into energy non-conserving and energy non-conserving effects. While the pseudo-diabatic PSSH approach can capture most of the exact results qualitatively (and often quantitatively), Tully's FSSH algorithm has large errors. For more benchmarking results and a further discussion of the phase-space adiabatic surfaces, see the SI.

\onecolumngrid 

\twocolumngrid

\paragraph{Discussion and Perspective.}
The present results with pseudo-diabatic PSSH have demonstrated a surprising degree of accuracy by successfully incorporating both nonadiabatic effects and Berry curvature effects. And yet, interestingly, the entire concept of Berry force has been replaced: we no longer apply a pseudo-magnetic field to motion along an adiabat, but rather use the relevant Hamiltonian dynamics as applicable to a magnetic field. Thus, one must presume that the present approach would be optimal for running surface hopping in an external magnetic field as well. By using phase-space adiabatic surface hopping, it would appear that one can capture very new physics (all while reducing to normal FSSH when a $2 \times 2$ Hamiltonian is real-valued). In this same spirit, other semiclassical approaches, e.g. multiple spawning, might also benefit by employing a pseudo-diabatic representation and running along phase-space adiabats whenever one encounters complex-valued Hamiltonians. 
More generally, we are confident that the pseudo-diabatic PSSH algorithm proposed here (or some version thereof) is the optimal framework for semiclassical simulation of large, complicated nonadiabatic systems where electronic spin effects are important.

Now, in making the claim above, our confidence is based on several factors. First, over the past few years, our research group has worked investigate many different FSSH algorithms (incorporating Berry curvature effects) within a host of two-dimensional models Ref.~\cite{wu2021Semiclassical}. We found that for many problems, if one chooses the right rescaling approach, FSSH can yield good results; however, the final algorithm \cite{wu2021Semiclassical} always felt overly complicated. By contrast, the present PSSH algorithm is simple to understand and to implement.
Second, the algorithm in Ref.~\cite{wu2021Semiclassical} fails when the diabatic coupling is very small; in such a case, the Berry force is not important and should not play a role in FSSH; the present PSSH algorithm does not fail in this limit. See Fig.~S4 in the SI.
Third, the algorithm in Ref.~\cite{wu2021Semiclassical} also fails when $W$ gets large (even though, one might presume that the Berry force grows larger and larger). This failure is completely corrected by the present PSSH approach. See Fig.~S5 in the SI. In short, the PSSH \ansatz appears to be the optimal approach moving forward; in the future, it might be best to refer not to Berry forces per se but rather to nonadiabatic dynamics in phase space.

Looking forward, our initial success here would appear to be only the first step in a long road towards running on-the-fly nonadiabatic dynamics with nuclei, electrons and spin. There are many obstacles that must be addressed and/or overcome. Here, we will list a few (though the list is not exhaustive).
First, the success of our algorithm relies on the premise that there is an intrinsic diabatic basis to dress (as in Eq.~\eqref{eq:hel}). \footnote{In a future publication, we will address what are the results if we choose a different pseudo-diabatic basis.}
How should we select such an optimal basis in practice? For an idealized, well-defined avoided crossing problem as in Fig.~\ref{fig:crossing}, one can guess the correct proper diabats almost intuitively. However, for systems with a complicated topology, e.g. a conical intersection or a crossing between a singlet and a set of triplets \cite{bian2021Modeling}, picking the correct diabats would appear much more difficult. Semiclassical dynamics can be very sensitive to the choice of a diabatic basis, and a systematic understanding of the impact of diabatization (as well as practical algorithms for choosing diabats) is essential.

At this point, it is worthwhile to compare and contrast our approach with Shenvi's adiabatic PSSH algorithm \cite{shenvi2009Phasespace}. It is known for real-valued Hamiltonians that the algorithm often performs better than Tully's FSSH (in a position-adiabat basis) -- at least in the adiabatic regime \cite{shenvi2009Phasespace,gherib2016inclusion}. While Shenvi's algorithm has so far not been applied previously to complex-valued Hamiltonians \footnote{We will address the performance of Shenvi's adiabatic phase-space surface hopping algorithm in a future publication.}, if one were to make such an attempt, one would necessarily need to choose a gauge for the adiabats (before diagonalizing into a superadiabatic basis). In other words, our present need for a good diabatic basis would correspond to the need for a good gauge within Shenvi's adiabatic PSSH algorithm. There is no free lunch, but future work will need to run many simulations to make sure we find the most stable approximations. \footnote{In principle, one can anticipate another obstacle when applying Shenvi's algorithm to nonadiabatic dynamics with spin-orbit coupling: namely, the presence of degenerate states which makes it impossible to isolate unique adiabatic states. Nevertheless, in the future, it will be essential to benchmarks the Shenvi algorithm for complex-valued Hamiltonians.}

Second, the question of decoherence must be addressed and benchmarked. Within standard FSSH, decoherence appears to be very complicated for complex electronic Hamiltonians. After all, different Berry forces would appear to lead to wave packet separation in the vicinity of an avoided crossing \cite{wu2021Semiclassical} -- whereas, in the context of real-valued Hamiltonians, decoherence arises only {\em after} wavepackets leave the vicinity of a crossing. \cite{subotnik2016Understanding,fang1999Improvement,horenko2002Quantumclassical,subotnik2011Decoherence} Within PSSH, however, it would appear that this distinction is removed and decoherence again is simple -- wavepackets separate only after the packets leave the crossing region now as driven by a difference in adiabatic {\em phase-space eigenforces}. This hypothesis must be checked in the future. In the future, we will also need to address the question of velocity reversal, which is known to be important for many simulations with frustrated hops. \cite{subotnik2016Understanding,jasper2011NonBorn,muller1997Surfacehopping} See Fig.~S3 in the Supplementary Information for some preliminary data.

Third, for systems with more than two states and couplings between each pair of diabats, the construction of pseudo-diabats may be impossible if we insist on $(i)$ a one-to-one mapping between pseudo-diabats to diabats and $(ii)$ a strictly real-valued the electronic Hamiltonian. For example, consider the following diabatic electronic Hamiltonian:
\begin{align}
    h_{\text{el}} = \begin{bmatrix} h_1 & V_1e^{i\phi_1} & V_2e^{i\phi_2} \\ V_1e^{-i\phi_1} & h_2 & V_3e^{i\phi_3} \\ V_2e^{-i\phi_2} & V_3e^{-i\phi_3} & h_3 \end{bmatrix}
\end{align}
If $\phi_1,\phi_2,\phi_3$ are not related to each other, there is no choice of simple pseudo-diabats for making $h_{\text{el}}$ real-valued. In such a case, one will either need to accept a complex-valued pseudo-diabatic Hamiltonian or apply a more general ``pre-conditioning'' diabatization. Future research is clearly required on this front.

Fourth and finally, it is known that the surface hopping algorithm can be derived roughly from the mixed quantum-classical Liouville equation (QCLE) \cite{subotnik2013Can,kapral1999Mixeda} if one makes some very strong approximations--e.g. the single-trajectory approximation, etc. In this letter, upon hopping we have followed standard procedure\cite{tully1990Molecular, kapral1999Mixeda} and conserved energy by rescaling momentum. Nevertheless, according to Eq.~\eqref{eq:hop}, one might presume that the more rigorous framework is to rescale both position and momentum \cite{yang2009optimal} upon hopping. 
\footnote{For our current model Hamiltonian~\eqref{eq:hmodel}, one can justify using momentum rescaling alone (without position rescaling) to maintain energy conservation because $\dot{\mathbf{P}}\cdot\mathbf{d}_{jk}^P=-\nabla_R{E_j}\cdot\braket{\psi_j}{\nabla_P\psi_k} = 0$ (assuming the trajectory is on surface $j$). After all, $E_j$, $\ket{\psi_j},$ and $\ket{\psi_k}$ are all functions of only $x$ and $P_y$ (see the Supplementary Info for the derivation). That being said, this approach may not robust for a more general case.}
In the future, one will necessarily need to investigate the formal foundations of phase-space surface hopping (starting with the QCLE), and systematically analyze the rescaling approach. Ideally, one would also like to connect with multicomponent WKB theories as well.\cite{littlejohn1991Geometric,weigert1993Diagonalization}

In summary, we have proposed a pseudo-diabatic phase-space surface hopping (PSSH) for propagating complex-valued nonadiabatic dynamics for two-state avoided crossing problems. The approach is simple, intuitive, and with strong potential for broad applicability. In our test models, our method has achieved a reasonably high accuracy and correctly incorporated all Berry curvature effects (without directly applying a pseudo-magnetic field). Our results indicate that performing a basis transformation as well as using a phase-space basis are crucial when modeling nonadiabatic dynamics in complex-valued systems. Looking forward, we are very hopeful that this algorithm can be applied to larger, \abinitio systems spin-related phenomena, including chemical reactions displaying magnetic field effects \cite{hore2020Spina} and chiral induced spin separated dynamics. \cite{naaman2012ChiralInduced}

This material is based on the work supported by the National Science Foundation under Grant No. CHE-2102402.

\end{document}


\title{Supplementary Information for ``A Phase-Space Semiclassical Approach for Modeling Nonadiabatic Nuclear Dynamics with Electronic Spin''}
\author{Yanze Wu}
\email{wuyanze@sas.upenn.edu}
\affiliation{Department of Chemistry, University of Pennsylvania, Philadelphia, Pennsylvania 19104, USA}
\author{Xuezhi Bian}
\email{xzbian@sas.upenn.edu}
\affiliation{Department of Chemistry, University of Pennsylvania, Philadelphia, Pennsylvania 19104, USA}
\author{Jonathan Rawlinson}
\email{jonathan.rawlinson@manchester.ac.uk}
\affiliation{School of Mathematics, The University of Manchester, Oxford Rd, Manchester M13 9PL, United Kingdom}
\author{Robert G. Littlejohn}
\email{robert@wigner.berkeley.edu}
\affiliation{Department of Physics, University of California, 366 Physics North MC 7300, Berkeley, CA, 94720-7300}
\author{Joseph E. Subotnik}
\email{subotnik@sas.upenn.edu}
\affiliation{Department of Chemistry, University of Pennsylvania, Philadelphia, Pennsylvania 19104, USA}
\date{\today}

\maketitle

In this supplementary information file, we provide an analysis of the proper diabats corresponding to the Hamiltonian in Eq.~\eqref{eq:hmodel} of the paper. We also provide a few extra figures detailing some nuances of applying phase-space surface hopping (PSSH) to nonadiabatic dynamics problems with complex-valued Hamiltonians. 

\section{Choice of Basis of for the Nonadiabatic Hamiltonian in Eq.~(13)} \label{sec:basis}

As discussed in the main body of the paper, for PSSH we require a set of proper diabats that asymptotically coincide with adiabats outside the crossing; the pseudo-diabats are then chosen so as to make the Hamiltonian real-valued. Here we demonstrate why we must select such a set of basis functions (the proper diabats and pseudo-diabats) using the two-dimensional nonadiabatic Hamiltonian in Eq.~\eqref{eq:hmodel} as an example. We will show that selecting this basis can sometimes reduce a two-dimensional problem to an effectively one-dimensional form, which has a stable rescaling direction as well and a good, conserved quantum number. 


We rewrite the Hamiltonian~\eqref{eq:hmodel} here:
\begin{align} \tag{13} \label{eq:hmodel}
    h_{\text{el}}(x,y) = A \begin{bmatrix} -\cos{\theta(x)} & e^{iWy}\sin{\theta(x)} \\ e^{-iWy}\sin{\theta(x)} & \cos{\theta(x)} \end{bmatrix}
\end{align}
Note that this Hamiltonian is two-dimensional. For this model Hamiltonian, the diabats are already the `proper diabats', i.e. they coincide with adiabats outside the crossing. By choosing pseudo-diabats as $\ket{\xi_0}=\ket{\chi_0}$ and $\ket{\xi_1}=\ket{\chi_1}e^{-iWy}$, we arrive at the pseudo Born-Oppenheimer Hamiltonian:
\begin{align}
    H_{\text{PD}} = \frac{P_x^2}{2M}+\frac{1}{2M}\begin{bmatrix}P_y^2 & 0 \\ 0 & {(P_y-\hbar W)}^2 \end{bmatrix} + A \begin{bmatrix} -\cos{\theta(x)} & \sin{\theta(x)} \\ \sin{\theta(x)} & \cos{\theta(x)} \end{bmatrix}
    \label{eq:hpdmodel}
\end{align}
Note that Hamiltonian $H_{\text{PD}}$ has no explicit dependence on $y$, which means that quantum mechanically, operator $P_y$ is time independent:
\begin{align}
    \dot{P}_y = \frac{i}{\hbar}[H_{\text{PD}}, P_y] = 0
\end{align}
Therefore, Hamiltonian~\eqref{eq:hpdmodel} is essentially a one-dimensional Hamiltonian, and the phase-space momentum in the $y$ direction is conserved. Semiclassically, this indicates that the nonadiabatic vector is unambiguously along $x$. 

Now, if the initial diabats were chosen differently, such momentum conservation and stable rescaling direction does not arise. For example, if we were to choose $\ket{\chi_0'} = (\ket{\chi_0}+\ket{\chi_1})/\sqrt{2}, \ket{\chi_1'} = (\ket{\chi_0}-\ket{\chi_1})/\sqrt{2}$, then the electronic Hamiltonian becomes
\begin{align}
    h_{\text{el}}'(x,y) = A \begin{bmatrix} \sin{\theta(x)}\cos(Wy) & -(\cos{\theta(x)}+i\sin{\theta(x)}\sin(Wy)) \\ -(\cos{\theta(x)}-i\sin{\theta(x)}\sin(Wy)) & -\sin{\theta(x)}\cos(Wy) \end{bmatrix}
\end{align}
We can still force $h'_{\text{el}}$ to have the form of Eq.~(3) by defining $\theta'(x,y)=-\acos(\sin\theta\cos(Wy))$ and $\phi'(x,y)=\atan(\tan\theta\sin(Wy))$:
\begin{align}
    h_{\text{el}}'(x,y) = A \begin{bmatrix} -\cos{\theta'} & \sin{\theta'}e^{i\phi'} \\ \sin{\theta'}e^{-i\phi'} & \cos{\theta'} \end{bmatrix}
\end{align}
Thereafter, the corresponding pseudo Born-Oppenheimer Hamiltonian is still of the form:
\begin{align}
    H_{\text{PD}} = \frac{P_x^2}{2M}+\frac{1}{2M}\begin{bmatrix}P_y^2 & 0 \\ 0 & {(P_y-\hbar \nabla\phi')}^2 \end{bmatrix} + A \begin{bmatrix} -\cos{\theta'} & \sin{\theta'} \\ \sin{\theta'} & \cos{\theta'} \end{bmatrix}
    \label{eq:nonproperpd}
\end{align}
However, since both $\theta'$ and $\phi'$ are functions of $x,y$, the Hamiltonian $H_{\text{PD}}'$ is no longer one dimensional and $P_y$ is not conserved. Thus, the basis $\{\ket{\chi_0'},\ket{\chi_1'}\}$ does not reveal the underlying symmetry for Hamiltonian~\eqref{eq:hmodel}. Moreover, the semiclassical propagation using Eq.~\eqref{eq:nonproperpd} becomes unstable, since $\nabla\theta'$ is now rotating in the $x-y$ plane. Thus, for PSSH, it is imperative to find a good set (i.e. a proper set) of diabatic states.

\section{Features of Nonadiabatic Dynamics with Complex-valued Hamiltonians} \label{sec:feature}

In Fig.~2 of the main text, we have shown typical scattering results for the model Hamiltonian~\eqref{eq:hmodel}. Here we provide a detailed analysis of the dynamics. In short, there are two typical features of scattering with Hamiltonian~\eqref{eq:hmodel}: The kinetic momentum shift in the $y-$direction and reflection at low momenta. We will now show how these features arise and how they are connected to each other.

The phase-space adiabatic energies can be calculated by diagonalizing $H_{\text{PD}}$:
\begin{align}
    E_{\pm}(P_x,P_y,x) = \frac{P_x^2}{2M} + \frac{P_y^2}{2M} + \frac{g}{2} \pm \sqrt{A^2+\frac{g^2}{4}+Ag\cos{\theta}}
    \label{eq:epdmodel}
\end{align}
where
\begin{align}
    g = -\frac{\hbar WP_y}{2M} + \frac{\hbar^2W^2}{2M} = \frac{{(P_y-\hbar W)}^2-P_y^2}{2M}
    \label{eq:eg}
\end{align}
Note that $g$ is the energy required to shift $P_y$ to $P_y-\hbar W$. 

Using Eqs.~(9),(10)) and the fact that $\mathrm{d}g/\mathrm{d}P_y = -\hbar W/2M$, we can find the the semiclassical equation of motions for $y$ and $P_y$:
\begin{align}
    \dot{P_y} &= 0 \label{eq:py0}\\ \label{eq:py}
    \dot{y} &= \begin{cases}
        \frac{P_y}{M}-\frac{\hbar W}{2M}(1+ \frac{\frac{g}2+A\cos{\theta}}{\sqrt{A^2+\frac{g^2}{4}+Ag\cos{\theta}}}), & \text{on surface }\ket{+}, \\
    \frac{P_y}{M}-\frac{\hbar W}{2M}(1- \frac{\frac{g}2+A\cos{\theta}}{\sqrt{A^2+\frac{g^2}{4}+Ag\cos{\theta}}}), & \text{on surface }\ket{-}
    \end{cases}    
\end{align}

\begin{figure}[H]
    \begin{center}
        \includegraphics[width=1\columnwidth]{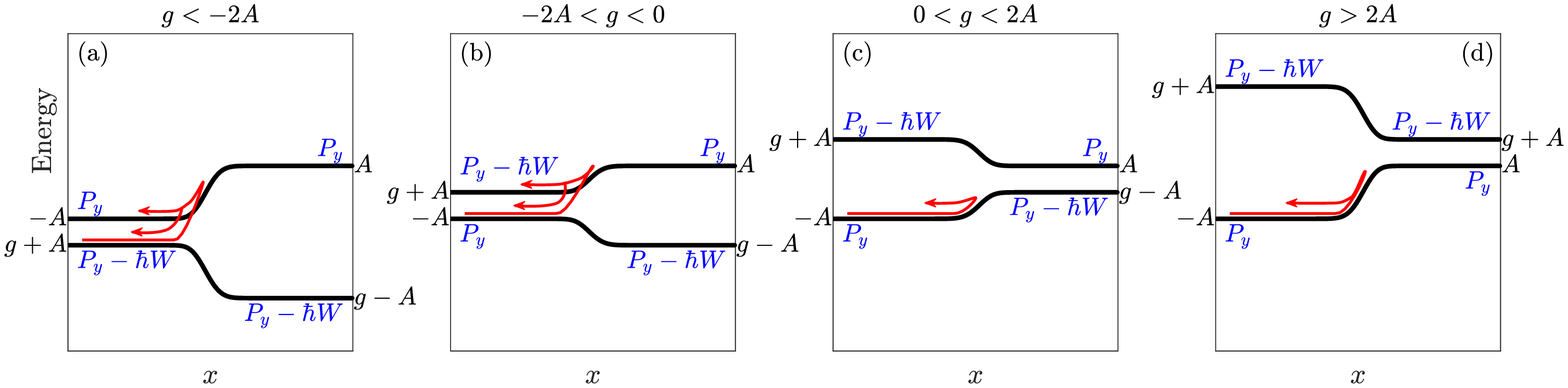}
    \end{center}
    \caption{Schematic figure of the energy (shifted by $-P_x^2/2M-P_y^2/2M$) of the phase-space adiabats as a function of $x$. The values of asymptotic kinetic momentum $P^k_y$ are marked as blue texts, and the paths for reflection of a wavepacket incoming from $\ket{-}$ are marked in red arrows. Note that the energy of phase-space adiabat $\ket{-}$ is always below the energy of phase-space adiabat $\ket{+}$. }\label{fig:psadiabat}
\end{figure}

According to Eq.~\eqref{eq:py0}, we see that $P_y$ is a constant. However, according to
Eq.~\eqref{eq:py}, the kinetic momentum $P^k_y \equiv M\dot{y}$ is a function of $x$. Moreover, depending on the value of $g$, the adiabatic energies and asymptotic momenta have different limits. For instance, note that as $x\to\pm\infty$, $\cos{\theta}\to\pm 1$. Thus, asymptotically, 
\begin{align}
\frac{g}{2} \pm \sqrt{A^2+\frac{g^2}{4}+Ag\cos{\theta}}\to\frac{g}{2} \pm \sqrt{A^2+\frac{g^2}{4} \pm Ag}
\end{align}
and 
\begin{align}
    1\pm\frac{\frac{g}2+A\cos{\theta}}{\sqrt{A^2+\frac{g^2}{4}+Ag\cos{\theta}}}\to
    1\pm\frac{\frac{g}2 \pm A}{\sqrt{A^2+\frac{g^2}{4}\pm Ag}}=1\pm\sgn(\frac{g}2\pm A)
\end{align}
When $x \to \pm \infty$, the value of $P^k_y$ becomes either $P_y$ or $P_y-\hbar W$; the asymptotic energy can be only $\pm A$ or $g \pm A$. For the exact details, see Fig.~\ref{fig:psadiabat}.

If one works out all of the cases in Fig.~\ref{fig:psadiabat}, it follows that exactly for those asymptotes where $P^k_y=P_y-\hbar W$, there is also an asymptotic shift $g$ in the energy. In other words, $P^k_y$ reflects an effective ``geometric barrier''. These results agree with our previous observations and explanations using a Berry curvature argument. \cite{miao2019extension,subotnik2019demonstration}.
From Fig.~\ref{fig:psadiabat}, we have the capacity to gain a very new understanding of the resulting physics. To do so, imagine a trajectory incoming from the phase-space adiabat $\ket{-}$ from the left side. (The electronic character of the phase-space adiabat will be discussed in Sec.~\ref{sec:electronic} below.) From energetic considerations alone, one can use Fig.~\ref{fig:psadiabat} to determine just when a trajectory will reflect (as demonstrated with red arrows in Fig.~\ref{fig:psadiabat}). There are four relevant cases:

\begin{enumerate}
\item $g>0$, and $P_x^2/2M < \min(2A, g)$. Here the trajectory is reflected on the same phase-space adiabat (Fig.~\ref{fig:psadiabat}c, \ref{fig:psadiabat}d) as it approaches on. In the main body of the paper, this is the case where $W=-5$ and the wavepacket incomes from diabat 0. The reflection condition dictates that $P_{\text{init}}<11.0$, which agrees with Fig.~2j.
\item $-2A<g<0$, and $2A+g<P_x^2/2M<2A$. Here the trajectory is reflected on the opposite phase-space adiabat (Fig.~\ref{fig:psadiabat}b), $\ket{+}$, and can emerge on either $\ket{+}$ or $\ket{-}$. In the main body of the paper, this is the case where $W=5$ and the incoming wavepacket lies along diabat 0. This condition gives $8.0<P_{\text{init}}<11.0$, which agrees with Fig.~2i and Fig.~2m.
\item $g<-2A$, and $-2A-g<P_x^2/2M<-g$. Here the trajectory is also reflected on the opposite phase-space adiabat (Fig.~\ref{fig:psadiabat}a), $\ket{+}$, and can emerge on either $\ket{+}$ or $\ket{-}$. In the main body of the paper, this is the case where $W=5$ and the incoming wavepacket is along diabat 1. This conditions gives $P_{\text{init}}<12.1$, which agrees with Fig.~2k and Fig.~2o.
\end{enumerate}

\section{Electronic Character of the Phase Space Adiabats} \label{sec:electronic}
\begin{figure}[H]
    \begin{center}
    \subfloat{\includegraphics[width=0.9\columnwidth]{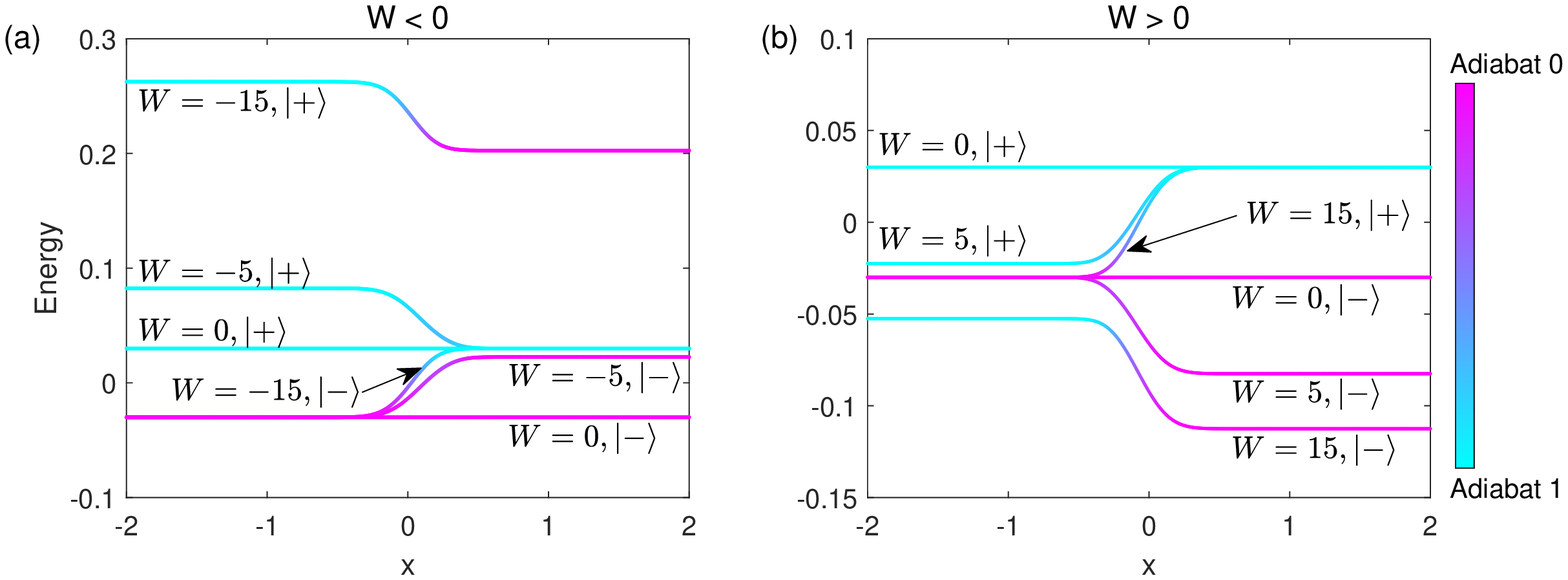}}
    
    \subfloat{\includegraphics[width=0.9\columnwidth]{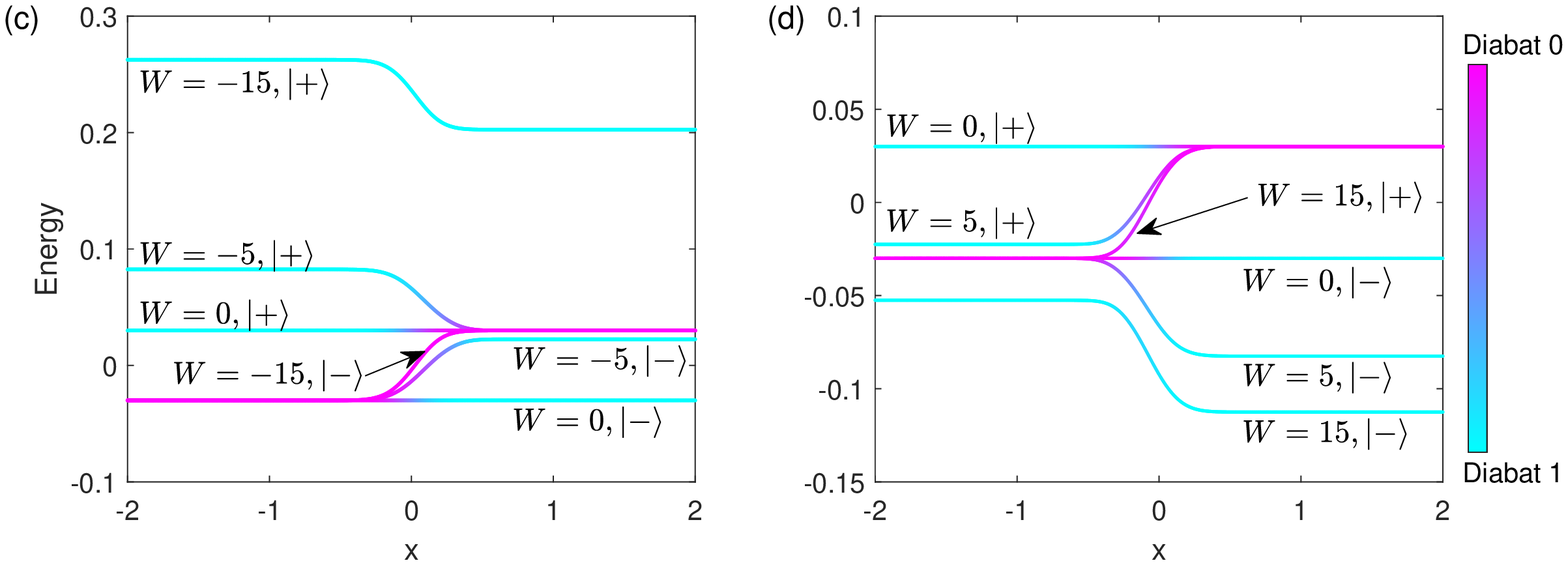}}
    \end{center}
    \caption{The energy surfaces of the phase-space adiabats (shifted by $-P_x^2/2M - P_y^2/2M$) with different values of $W$. The color of energy surface represents their overlap with position-space adiabats (for subfigures (a) and (c)) or diabats (for subfigures (b) and (d)). When $W=0$ or $W=\pm 5$, the phase-space adiabats can be mapped in a one-to-one fashion to the position-space adiabats. However, when $W=\pm 15$, the phase-space adiabats more closely resemble the diabats; the overlap with the corresponding position-space adiabat switches in the crossing region. Parameters are $A=0.03$, $B=3$ and $P_y^k=8$.}\label{fig:largew}
\end{figure}

The asymptotic kinetic momenta features in Fig.~\ref{fig:psadiabat} arise from the changing electronic character of the phase-space adiabats. To visualize this change in character, in Fig.~\ref{fig:largew} we plot both the phase-space adiabat energy {\em and} the overlap of the phase-space adiabats with the position-space adiabats and position-space diabats.

For the parameters used in our paper, we find that when the absolute value of $W$ is small, e.g. $P_y=8$ and $W = \pm 5$ (i.e. $\abs{g}<2A$), the phase-space adiabatic wave functions are effectively equivalent to the position-space adiabats. However, when the absolute value of $W$ is big (e.g. $W=\pm 15$ (i.e. $\abs{g}>2A$)), the character of the phase-space adiabats becomes inverted on the $x>0$ side: the character of the states changes. Thus, for large $W$, the phase-space adiabats more closely resemble the original diabats (not the adiabats).
Mathematically, the relative value between $\abs{g}$ and $2A$ is the critical threshold: for $\abs{g}<2A$, the character of the phase-space adiabats cross a function of $x$; for $\abs{g}>2A$, the character of the phase-space adiabats does not cross.

\section{Velocity Reversal}
\begin{figure}[H]
    \begin{center}
    \includegraphics[width=0.8\columnwidth]{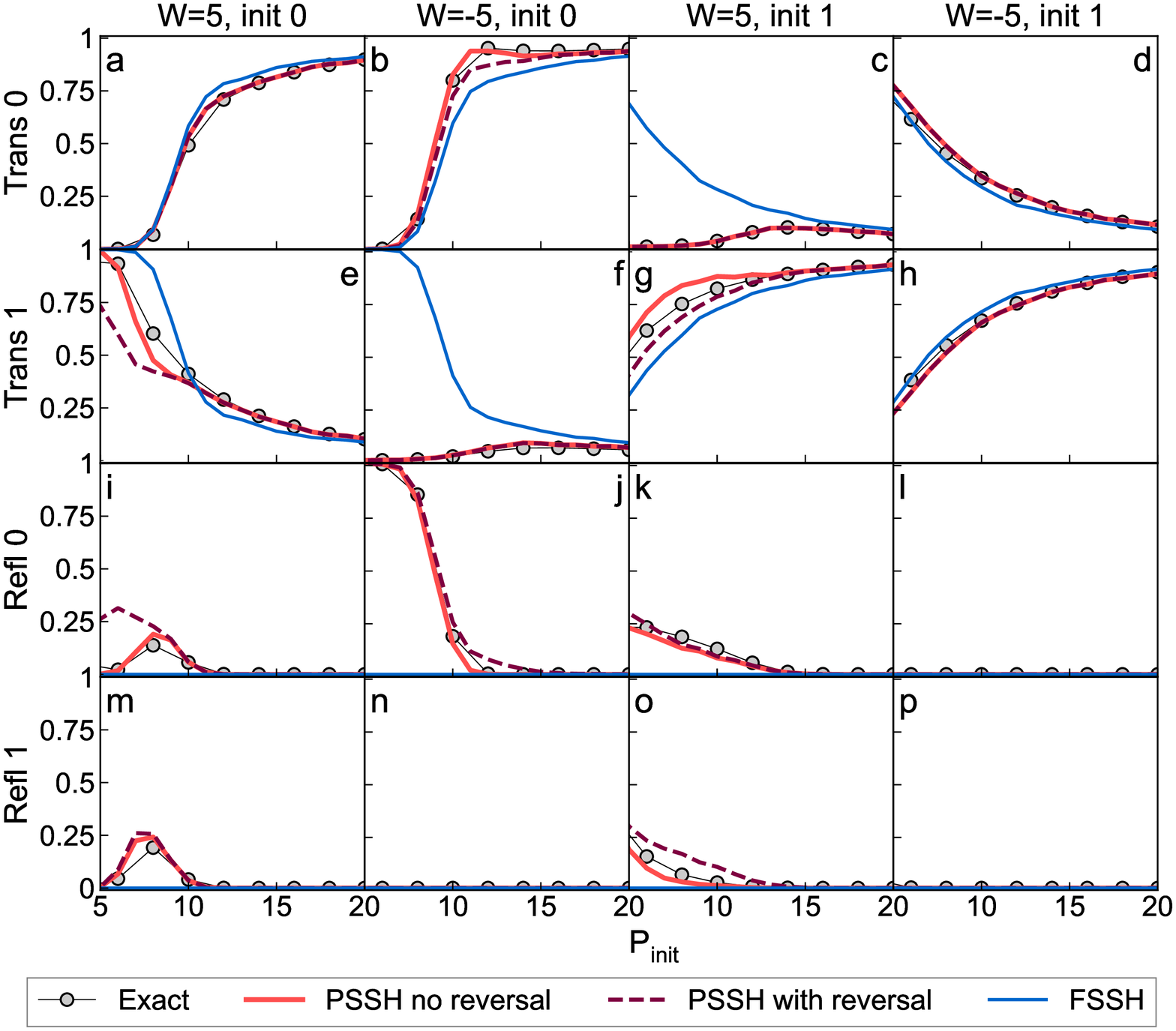}
    \end{center}
    \caption{State-to-state transmitted and reflected probabilities according to an exact wavepacket simulation, Tully's FSSH, and pseudo-diabatic PSSH with or without the velocity reversal method described in Ref.~\cite{jasper2011NonBorn}. The parameters are same as in Fig.~2, except that now $A=0.02$. Note that the overall impact of velocity reversal is small; however, velocity reversal can sometimes cause unphysical reflection at low momenta (e.g. subplot (i)).}\label{fig:rev}
\end{figure}

One important nuance of surface hopping is the question of if and when to reverse velocities after a frustrated hop. To that end, below we compare the performance of PSSH with or without velocity reversal following a frustrated a hop. Here we used the ``grad V'' velocity reversal rule proposed by Jasper {\em et al} \cite{jasper2011NonBorn}; in other words, when a hop from surface $j$ towards surface $k$ fails, the trajectory's momentum is reversed if $(\mathbf{P}\cdot\mathbf{d}_{jk})(\nabla E_k\cdot\mathbf{d}_{jk})>0$. In preliminary tests, we have found that the impact of velocity reversal is largest when $A$ is around 0.02; therefore we choose $A=0.02$ for our test case. Results are shown in Fig.~\ref{fig:rev}. The overall difference between PSSH with or without velocity reversal is small, except for rather low incoming momenta (e.g. $P_{\text{init}}<8$). Adding velocity reversal gives slightly better results for diabatic transmission probability when $W=5$ and $\chi_{\text{init}}=\chi_1$, as shown in Fig.~\ref{fig:rev}g. However, in other cases, e.g. $W=5$ and $\chi_0=0$, velocity reversal overestimates the reflected population (Fig.~\ref{fig:rev}i). 
These results suggests that the complex-valuedness of Hamiltonian may complicate the velocity reversal correction and will require more care in benchmarking the optimal approach. For simplicity, we have omitted all velocity reversal corrections in the main body of the paper.

\section{Comparison with Our Previous Rescaling Extension to FSSH}
\begin{figure}[H]
    \begin{center}
    \includegraphics[width=0.8\columnwidth]{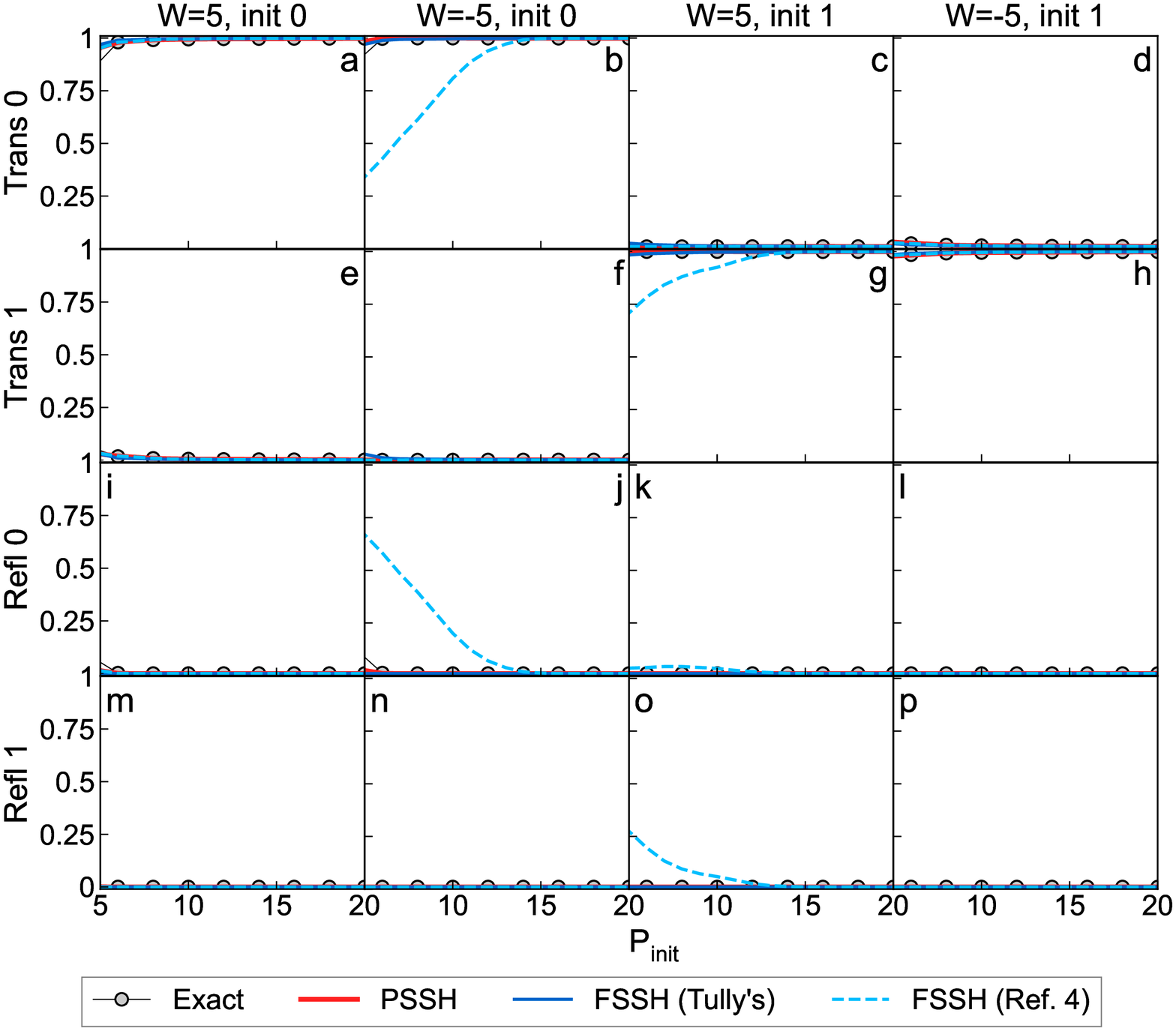}
    \end{center}
    \caption{State-to-state transmitted and reflected probabilities according to an exact wavepacket simulation, pseudo-diabatic PSSH, Tully's FSSH \cite{tully1990Molecular} and FSSH with Berry force (as proposed in Ref.~\cite{wu2021Semiclassical}). The parameters are same as in Fig.~2, except now $A=0.002$. Note that both Tully's FSSH and pseudo-diabatic PSSH can capture the correct population, but Ref.~\cite{wu2021Semiclassical} gives too much reflection.}\label{fig:a0.002}
\end{figure}

\begin{figure}[H]
    \begin{center}
    \includegraphics[width=0.8\columnwidth]{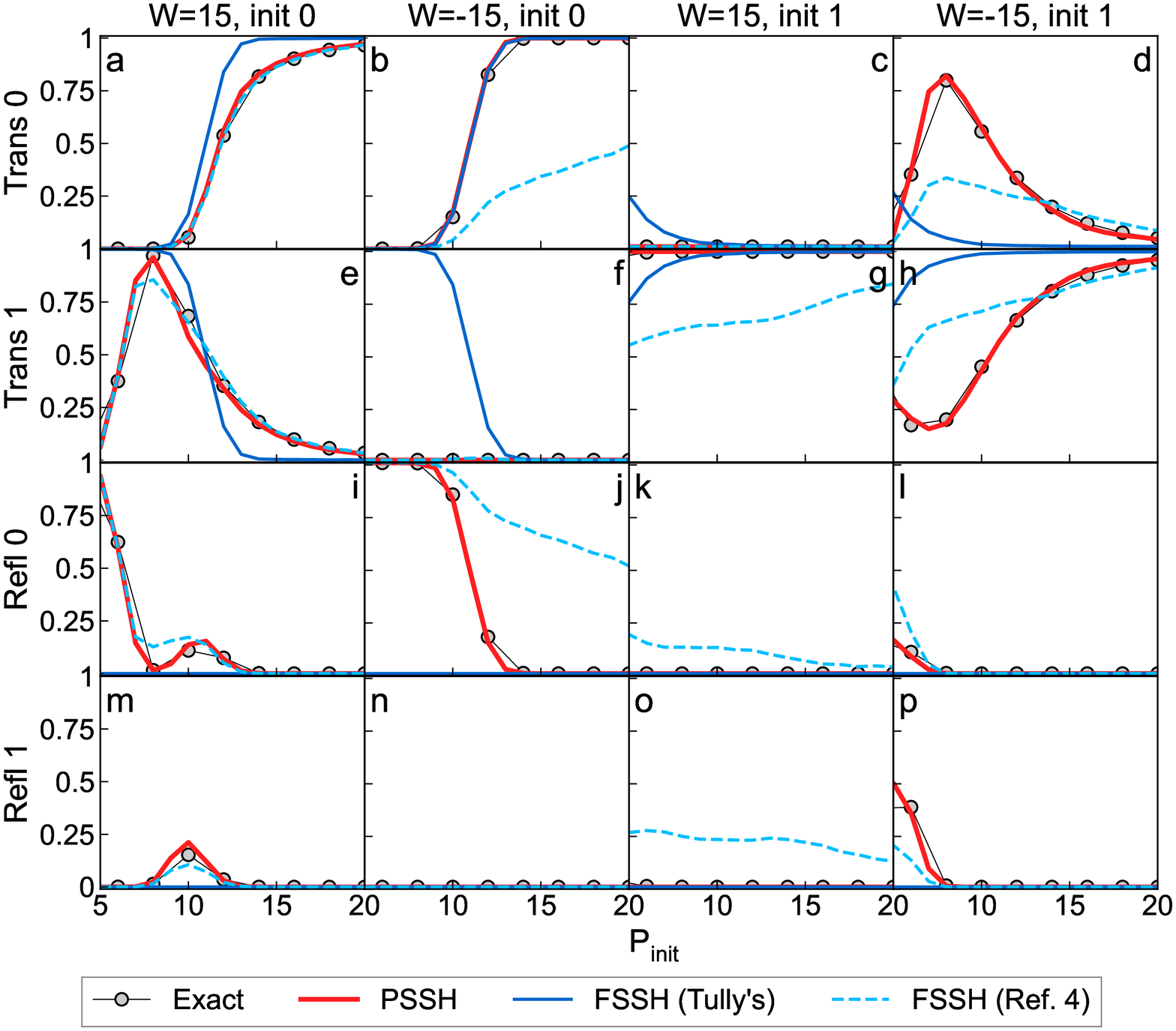}
    \end{center}
    \caption{Same as in Fig.~\ref{fig:a0.002}, except now $A=0.03$ and $W=\pm 15$. Note that FSSH (Tully style \cite{tully1990Molecular} or with Berry force \cite{wu2021Semiclassical}) fails for low incoming momentum, but PSSH works quite well.}\label{fig:w15}
\end{figure}

In order to evaluate the performance of the proposed PSSH approach, one of the most important benchmark is to compare PSSH versus our previous incarnation of FSSH (whereby we explicitly included a Berry force correction).
To that end, we below we compare results using Ref.~\cite{wu2021Semiclassical} versus our current algorithm in (1) the diabatic limit and (2) the case where the phase of the diabatic coupling oscillates very quickly in space. For case (1), we choose $A=0.002$, $W=\pm5$ (see Fig.~\ref{fig:a0.002}). For case (2), we choose $A=0.03$, $W=\pm 15$ (see Fig.~\ref{fig:w15}). From the data, we clearly see that pseudo-diabatic PSSH can handle both cases very well, while FSSH with Berry force (i.e. Ref.~\cite{wu2021Semiclassical}) fails.


To understand why PSSH performs so well in Fig.~\ref{fig:w15} (and why standard FSSH fails) when the absolute value of $W$ is large ($W=15$), consider Fig.~\ref{fig:largew}.
Imagine a wavepacket incoming from the upper position-space adiabatic surface. In the phase-space adiabatic picture, this initialization corresponds to initialization on state $\ket{-}$, the {\em lower} phase-space adiabat (See Fig.~\ref{fig:largew}b).
As discussed above, for large $W$, this phase-space adiabat state resembles a diabatic state as a function of $x$; there is no strong avoided crossing to the upper phase-space adiabat, $\ket{+}$. Therefore, the wavepacket should transmit exclusively on the $\ket{-}$ surface (which is indeed found by PSSH and exact quantum dynamics). However, standard FSSH, using the position-space adiabatic picture, still presumes that the wavepacket starts from the upper position-space surface and undergoes strong nonadiabatic interactions with the lower position-space adiabat. Thus, not surprisingly, standard FSSH erroneously predicts some transmission on both outgoing surfaces (see Fig.~\ref{fig:w15}c and Fig.~\ref{fig:w15}g). Vice versa, FSSH with Berry force (Ref.~\cite{wu2021Semiclassical}) predicts far too much reflection (Fig.~\ref{fig:w15}j).